\documentclass[12pt,reqno,a4paper]{article}
\usepackage{amsmath,amsfonts,amssymb,graphicx,epsfig,pict2e,rotating}
\usepackage[english]{babel}
\usepackage[noxcolor]{pstricks}
\usepackage{hyperref}
\hypersetup{backref,
colorlinks=true,
citecolor=blue}
\numberwithin{equation}{section}

\newcommand{\be}{\begin{equation}}
\newcommand{\ee}{\end{equation}}

\begin{document}
\setcounter{page}{1}

\vspace{8mm}
\begin{center}
{\Large {\bf Free Field realization of the $\hat{\mathcal{D}}_q$  Algebra for the $\eta$-$\xi$ system, Integrals of Motion and Characters   }}

\vspace{10mm}
 {\Large Alessandro Nigro\footnote{Email: Alessandro.Nigro@mi.infn.it}}\\
 [.4cm]
  
  \vspace{10mm}
 { Alessandro Nigro\\
 Dipartimento di Fisica and INFN- Sezione di Milano\\
 Universit\`a degli Studi di Milano I\\
 Via Celoria 16, I-20133 Milano, Italy\\
 Alessandro.Nigro@mi.infn.it}\\
 [.4cm]

  \end{center}

\vspace{8mm}
\centerline{{\bf{Abstract}}}
\vskip.4cm
\noindent
We introduce a free field realization of the central extension of the Lie algebra  $\mathcal{D}_q$  of  difference operators  on the circle in terms of the fermionic $\eta$-$\xi$ system. This realization admits a nontrivial Jordan block structure. We also review the free field realization of  $\mathcal{W}_{1+\infty}$ algebra, and point out some relations beween its generators of weight zero and the local integrals of motion of Bazhanov Lukyanov and Zamolodchikov. Finally we compute the finitized characters, and the continuum characters of the Local Integrals of Motion, and find out and interesting analogy with the generating functions for the counting of branched covers of elliptic curves. 

\renewcommand{\thefootnote}{\arabic{footnote}}
\setcounter{footnote}{0}
\section{Introduction}
In this paper we will study the 2 dimensional Conformal Field Theory corresponding to the $\eta$-$\xi$ \cite{kausch1}\cite{kausch2} ghost system. It is well known that this theory is a logarithmic CFT with central charge $c=-2$, this theory has a large number of nontrivial features. Firstly there is a $U(1)$ symmetry associated to the conservation of the ghost number. If one restricts himself to the "small" algebra generated by $\partial\xi$,$\eta$ this symmetry is enhanced to a $SL(2)$ symmetry. This symmetry is only global  and is related to the presence  of a $\mathcal{W}_3$ algebra.\\
In this work we will introduce a two new $q-$deformation of the $U(1)$ ghost current, this $q-$deformation will give rise to a free field realization of the central extension Lie algebra of  difference operators on the circle $\hat{\mathcal{D}}_q$   \cite{KR}. We will analyze in detail the commutation relations of this algebra, and by introducing the expansion of the generators $D_n(q)$ for $q\to 1$ with $q=e^{\hbar}$ and $\hbar\to 0$ one has that the order $\hbar^0$ reproduces the $U(1)$ current, while the the order $\hbar$ is proportional to some $\mathcal{L}_n$ which form a Virasoro algebra with  central charge $c=1$. Moreover it is well known \cite{KR} that the algebras $\hat{\mathcal{D}}_q$ admit a Virasoro subalgebra with continuously varying central charge. This is realized by adding to the stress energy tensor a total derivative of the $U(1)$ current \cite{ghostsyst}. As it is well known in other cases of deformed  Algebras, the higher orders in $\hbar$ will generate higher spin operators which are related to the Integrals of Motion of Bazhanov Lukyanov and Zamolodchikov \cite{blz}. We analyze in detail the relation between the higher orders in the $\hbar$ expansion and the BLZ integrals of Motion at $c=-2$ which have been obtained for Symplectic Fermions in \cite{nigropol}.\\
We also give a brief review of the free field realization of the$\mathcal{W}_{1+\infty}$ (also called  $\hat{\mathcal{D}}$) algebra in terms of the ghost system \cite{AFMO}\cite{M}, and show that the generators of weight zero  are related to the integrals of motion.\\
We finally analyze the characters of the deformed dilatation generator $D_0(q)$ by performing the trace over all eigenstates of the deformed dilatation operator. These eigenstates are well known to be labelled by two-column diagrams $\mathcal{D}$ and can be expressed in a simple way in terms of the modes of the ghost fields. We furthermore compute the finitized and continuum characters of all the IOM by performing the trace over all eigenstates of the IOM, and notably among them the character of the operator $W^{(0)}_0$ which is the zero mode of the primary spin $3$ field which together with other 2 spin $3$ fields forms the $\mathcal{W}$ algebra of the model. This character turns out to be related to the character of the $\mathcal{W}_3$ algebra \cite{characterW3}, and also related \cite{KZ} to the generating function of counting covers of an elliptic curve. \\
Moreover we compute the character of all the integrals of motion  which also turn out to be related to generating functions of counting covers of elliptic curves \cite{HO}. We also compute the character of an infinite linear combination of integrals of motion, which depends on infinite modular parameters and appeared before in \cite{HO}\cite{SBAO}, again as generating function of counting covers of  elliptic curves and as the character of the infinite wedge projective representation of the algebra of differential operators on the circle. These characters turn out to be  quasi modular forms \cite{KZ}\cite{SBAO}.

\section{The $\eta$-$\xi$ system}
The $\eta$-$\xi$  system is defined by the following stress energy tensor:
\be  T_2(z)=:\partial\xi(z)\eta(z): \ee
where the fields $\eta$-$\xi$ are defined by
\be \xi(z)=\sum_{n\in \mathbb{Z}}\frac{\xi_n}{z^n}  \ee
\be \eta(z)=\sum_{n\in \mathbb{Z}}\frac{\eta_n}{z^{n+1}}  \ee
and the modes satisfy the following anticommutation relations:
\be  \{\xi_m,\eta_n\}=\delta_{m+n,0} \ee
and $:\ldots:$ denotes fermionic normal ordering that means that the annihilation modes are moved to the right with a minus sign whenever two fields are interchanged. There is however an ambiguity in the normal ordering $:\xi_0\eta_0:$ which can be defined in general as:
\be :\xi_0\eta_0:=\lambda \eta_0\xi_0+(\lambda-1)\xi_0\eta_0 \ee
the convention $\lambda=\frac{1}{2}$ is employed in string theory, we shall however employ here the more convenient convention for $\lambda=1$.\\
The fields $\xi,\eta$ have the following correlation functions:
\be \Big<\eta(z)\xi(w)\Big>= \Big<\xi(z)\eta(w)\Big>=\frac{1}{z-w}  \ee
and
\be \Big<\eta(z)\eta(w)\Big>= \Big<\xi(z)\xi(w)\Big>=0  \ee
The $sl(2,\mathbb{C})$ invariant vacuum is characterized  by $\xi_n\Omega=\eta_n\Omega=0$ for $n>0$.\\
This theory admits an $U(1)$ symmetry which is generated by the following current:
\be J(z)=T_1(z)=:\xi(z)\eta(z): \ee
this current however does not define a well defined primary operator, as it can be seen by the following operator product expansion:
\be T_2(z)J(w)=\frac{-1}{(z-w)^3}+\frac{J(w)}{(z-w)^2}+\frac{\partial J(w)}{(z-w)}  \ee
from the operator product expansion of the stress energy tensor with itself 
\be  T_2(z)T_2(w)=\frac{-1}{(z-w)^4}+\frac{2T_2(w)}{(z-w)^2}+\frac{\partial T_2(w)}{(z-w)} \ee
one easily sees that the modes of the expansion
\be T_2(z)=\sum_{n\in\mathbb{Z}}\frac{L_n}{z^{n+2}}  \ee
satisfy the following Virasoro Algebra with central charge $c=-2$:
\be  [L_n,L_m]=(n-m)L_{n+m}-\frac{1}{6}n(n^2-1)\delta_{n+m,0} \ee
The zero mode $\xi_0$ can act on the vacuum ,  one has that there are 2 vacua $\Omega$ and $\xi_0\Omega$, and in general for each state $v$ there is always another state $\xi_0v$.\\
It is also possible to define twisted boundary conditions for the fields
\be \xi(e^{2\pi i}z)=-\xi(z) \ee
\be \eta(e^{2\pi i}z)=-\eta(z)  \ee
Where
\be \xi(z)=\sum_{n\in \mathbb{Z}}\frac{\xi_{n-\frac{1}{2}}}{z^{n-\frac{1}{2}}}  \ee
\be \eta(z)=\sum_{n\in \mathbb{Z}}\frac{\eta_{n+\frac{1}{2}}}{z^{n+1+\frac{1}{2}}}  \ee
In this case there are no zero modes and the twisted ground state $\mu$ is nondegenerate.\\
In \cite{FMS} it was observed that in the untwisted case the $\eta$-$\xi$ system  contains a "small" algebra generated by $\partial\xi$ and $\eta$, this algebra does not contain the mode $\xi_0$ and is characterized as the kernel of $\eta_0$ on the space of states of the $\eta$-$\xi$ system.\\
The states in the small algebra are of the following form
\be \xi_{-n_1}\ldots\xi_{-n_l}\eta_{-m_1}\ldots\eta_{-m_r}\Omega \ee
and we will denote them as $\big|\mathcal{D}\big>$ where $\mathcal{D}$ is the two column diagram $\mathcal{D}=(n_1,\ldots,n_l,m_1,\ldots,m_r)$ and the subscripts $l,r$ stand for left and right.\\
in the case of twisted boundary conditions one has the following generic state
\be \xi_{-n_1}\ldots\xi_{-n_l}\eta_{-m_1}\ldots\eta_{-m_r}\mu \ee
which we will also label by a two column diagram $\mathcal{D}$ and denote also as  $\big|\mathcal{D}\big>$ since there is basicly no source of ambiguity since the 2 sectors always appear in separate contexts.\\
On the "small" algebra the original $U(1)$ symmetry of the theory is enhanced to an $sl(2)$ symmetry.\\
By defining 
\be  \psi^+(z)=\eta(z)  \ee
\be \psi^-(z)=\partial\xi(z) \ee
the free fermion field $\vec{\psi}$ transforms as a $j=\frac{1}{2}$ representation of $sl(2)$\cite{kausch1}:
\be [J^+,J^-]=2J^0  \ee
\be [J^{\pm},J^0]=\pm J^0 \ee
\be [J^0,\chi^{\pm}(z)]=\pm\frac{1}{2}\chi^{\pm}(z) \ee
\be [J^{\pm},\chi^{\pm}(z)]=0 \ee
\be [J^{\pm},\chi^{\mp}(z)]=\chi^{\pm}(z) \ee
\be J^0\Omega=J^{\pm}\Omega=0 \ee
by virtue of this global symmetry the highest weight states will always fall into irreducible representations of $sl(2)$ carrying isospin $j\in\frac{1}{2}\mathbb{N}$:
\be \big|j,m\big>=\psi_{-2j}^{(+}\ldots\psi_{-j+m}^{+}\psi^{-}_{-j+m+1}\ldots\psi_{-1}^{-)}\Omega \ee
The fields with half integral isospin are fermionic whereas those with integral isospin are bosonic.\\
The bosonic sector is generated by the isospin 1 fields:
\be  W^{(+)}(z)= :\partial\eta(z)\eta(z):\ee
\be  W^{(0)}(z)=\frac{1}{2}(:\partial\eta(z)\partial\xi(z):+:\partial^2\xi(z)\eta(z):) \ee
\be W^{(-)}(z)= :\partial^2\xi(z)\partial\xi(z): \ee
The fields $W^{(i)}(z)$ are primary of spin $3$:
\be T_2(z)W^{(i)}(w)=\frac{3W^{(i)}(w)}{(z-w)^2}+\frac{\partial W^{(i)}(w)}{(z-w)}  \ee
and they form a $\mathcal{W}_3$ algebra.\\
 We notice in passing that the field $W^{(0)}(z)$ can be rewritten (for later convenience) as:
\be W^{(0)}(z)=\frac{1}{2}(-2T_3(z)+\partial T_2(z)) \ee
where we have introduced:
\be T_{2n}(z)=\partial^{n}\xi(z)\partial^{n-1}\eta(z) \ee
\be T_{2n-1}(z)=\partial^{n-1}\xi(z)\partial^{n-1}\eta(z) \ee
which will come in handy in the following sections.\\
We notice in passing that the mode of weight zero of the field $W^{(0)}(z)$ is expressed in terms of the modes of the fermi fields  as:
\be W^{(0)}_0= \sum_{m=1}^\infty m^2(N^+_m-N^-_m) \ee
where we have introduced the following number operators:
\be N^+_n=\xi_{-n}\eta_n  \ee
\be N^-_n=\eta_{-n}\xi_n  \ee
which for $m>0, k<0$ satisfy
\be [N^+_m,\xi_k]=\xi_k \delta_{m+k,0} \ee
\be [N^-_m,\eta_k]=\eta_k \delta_{m+k,0} \ee
\be [N^+_m,\eta_k]=0 \ee
\be [N^-_m,\xi_k]=0 \ee
we now introduce the following quantities
\be  I_k=\sum_{m=1}^\infty m^k(N^+_m+(-1)^{k+1}N^-_m)   \ee
which we shall call the Integrals of Motion, it is straightforward to see that their eigenvalues are given by
\be  I_k\big|\mathcal{D}\big>=(\sum_{m\in\mathcal{P}_1}m^k+(-1)^{k+1}\sum_{m\in\mathcal{P}_2}m^k) \big|\mathcal{D}\big>  \ee
where we have used the notation $\mathcal{D}=(\mathcal{P}_1|\mathcal{P}_2)$. \\
it is straightforward to see that the $I_{2k-1}$ have as eigenvalues the eigenvalues of the local integrals of motion of \cite{blz}, already obtained in \cite{nigropol}.\\
It is also straightforward to verify that they are in involution:
\be  [I_k,I_m]=0 \ee
We shall have to comment on the integrals of motion in the next sections, especially on their relations to the $\mathcal{W}_{1+\infty}$  and the $\hat{\mathcal{D}}_q$ algebras.\\
For now it suffices to notice that:
 \be W^{(0)}_0= I_2 \ee

\section{The $\hat{\mathcal{D}}$  and  $\hat{\mathcal{D}}_q$  Algebras}
The quantum  $\mathcal{W}_{1+\infty}$ \cite{M}\cite{AFMO}, also denoted by $\hat{\mathcal{D}}$ \cite{KR} is the quantum version of the $w_{1+\infty}$ algebra generated by the polynomials in the variables $z$ and $D=z\partial_z$. Given any regular function $f$ we may write its associated generator $z^m f(D)$. Let us denote the quantum generators of this algebra as $W(z^m f(D))$, the algebra associated to these generators is given by:
\be\begin{split} [W(z^m f(D)),W(z^n f(D))]=&W(z^{m+n} f(D+s)g(D))-W(z^{m+n} f(D)g(D+r))+\\&+c \sum_{j=1}^m f(-j)g(m-j) \delta_{n+m,0}   \end{split}\ee
We also introduce the generators $V^n_k=W(z^k D^n)$, which satisfy the following algebra:
\be  [V^n_r,V^m_s]=\sum_{k=0}^n \binom{n}{k}s^{n-k}V^{k+m}_{r+s} -\sum_{k=0}^m \binom{m}{k}r^{m-k}V^{k+n}_{r+s}+c \delta_{r+s}\sum_{j=1}^r (-j)^n (r-j)^m  \ee
This algebra can be realized in terms of the $\eta$-$\xi$ system as:
\be V^n_r=\oint\frac{dz}{2\pi i}z^r :D^n\xi(z)\eta(z): \ee
The commutation relations can be proven explicitly by computing the Operator Product Expansion of the fields by using the wick rules and computing a double contour integral, as it is described in the next section.\\
It is worth remarking that the modes of weight zero $V^n_0$, by virtue of the identity:
\be D^k(z^{-n})=(-1)^k n^k z^{-n} \ee
can be written as:
\be V^n_0= I_n  \ee
So that the subalgebra of modes of weight zero of $\mathcal{W}_{1+\infty}$ in this free field realization is generated by the integrals of motion, we shall comment on characters of this subalgebra in the last section.\\
We now turn our attention to the algebra $\mathcal{D}_q$ \cite{KR}, which is defined as the Lie algebra of difference operators on the circle. We first introduce the difference operator $D_q$:
\be  D_qf(z)=\frac{f(qz)-f(z)}{q-1}  \ee
We now introduce the following generators:
\be T_{m,n}=q^{(m+1)n}((q-1)D_q+1)^n  \ee
It follows, that upon parametrizing $q=e^{\hbar}$that these generators satisfy the following algebra:
\be [T_{m,n},T_{m',n'}]=2\sinh(\hbar(m'n-mn'))T_{m+m',n+n'}  \ee
 The algebra $\hat{\mathcal{D}}_q$ is defined as the central extension of this algebra, which is given by:
 \be  [T_{m,n},T_{m',n'}]=2\sinh(\hbar(m'n-mn'))T_{m+m',n+n'}  +c\frac{\sinh(\hbar m(n+n'))}{\sinh(\hbar(n+n'))}\delta_{m+m',0}  \ee
In the next section we shall introduce a free field realization of $\hat{\mathcal{D}}_q$  which up to now did not appear in the literature.\\

\section{The Free Field Realization of the Algebra $\hat{\mathcal{D}}_q$}
We start this section by introducing a new q-deformed current, this current is a deformation of the natural $U(1)$ current, but as we shall see it gives rise to a free field realization of the algebra $\hat{\mathcal{D}}_q$, namely the central extension of the Lie Algebra of finite difference operators on the circle.\\
We thus introduce the following spin-1 current:
\be D(z,q)=:\xi(q^{-1}z)\eta(q z): \ee
it can be seen by working out the calculation with Wick«s rules that this current has the following OPE with itself:
\be  D(z,q) D(w,t)=-\frac{q D(qw,qt)}{(z-qt w)}+\frac{ D(q^{-1}w,qt)}{q(z-(tq)^{-1}w)}+\frac{1}{(z-qt w)(z-(tq)^{-1}w)}    \ee 
now by introducing the following mode expansion
\be D(z,q)=\sum_{l\in\mathbb{Z}}\frac{D_l(q)}{z^{l+1}} \ee
one defines the commutator of two modes by the following contour integral:
\be [D_n(q),D_m(t)]=\oint\frac{dw}{2\pi i}\oint_{C_w}\frac{dz}{2\pi i}z^nw^m D(z,q)D(w,t)  \ee
where the contour $C_w$ encircles all poles of the OPE.\\
It is easy to see that the modes of the deformed current satisfy the following commutation relations
\be [D_n(q),D_m(t)]=(q^{ m}t^{-n}-q^{-m}t^{ n})D_{m+n}(qt)+\Bigg(\frac{(qt)^n-(tq)^{-n}}{qt-(tq)^{-1}}\Bigg)\delta_{n+m,0} \ee
We notice that for $q=e^{\gamma\hbar}, t=e^{\delta\hbar}$ this algebra reduces to the well known \cite{KR} central extension of the algebra $\hat{\mathcal{D}}_q$ of difference operators on the circle:
\be [D_n(e^{\gamma\hbar}),D_m(e^{\delta\hbar})]=2\sinh(\hbar(\gamma m-\delta n))D_{m+n}(e^{(\gamma+\delta)\hbar})+\frac{\sinh(\hbar(n(\gamma+\delta)))}{\sinh(\hbar(\gamma+\delta))}\delta_{n+m,0} \ee
Where however $\gamma, \delta$ are not restricted to be integers, so this algebra is somewhat more general.\\
If we now consider the expansion for $q\to1$ with $q=e^{\hbar}$ one has:
\be D(z,q)\sim T_1(z)-2z(T_2(z)-\frac{1}{2}\partial T_1(z) )\hbar+\ldots \ee
it is straightforward to see that this expansion can be expressed in closed form in terms of the $T_n(z)$ fields introduced in section 1. We may view these fields as integrals of motion, and suitable linear combinations of the $T_n(z)$ with higher derivatives of the lower spin fields will give rise to the local integrals of motion of \cite{blz} as we shall see.\\
Moreover we can introduce the analogous expansion for the modes of the q-deformed current:
\be  D_n(q)=\sum_{m=0}^\infty \hbar^m D_n^{(m)}  \ee
in particular one has that
\be D_n^{(1)}=-2\mathcal{L}_n \ee
where $\mathcal{L}_n$ satisfy the Virasoro algebra with central charge $c=1$.\\
Moreover the combination
\be  \mathcal{L}_n(\alpha)=L_n+(n+1)\frac{\alpha}{2}J_n   \ee
forms a Virasoro Algebra 
\be [\mathcal{L}_n(\alpha),\mathcal{L}_m(\alpha)]=(n-m)\mathcal{L}_{n+m}(\alpha)+\frac{c(\alpha)}{12}n(n^2-1)\delta_{n+m,0}  \ee
with central charge
\be  c(\alpha)=-3\alpha^2+6\alpha-2\ee
This Virasoro algebra is always contained as a subalgebra $\hat{\mathcal{D}}_q$.\\
It is worth pointing out, that for $c=1$ which corresponds to $\alpha=2$ the $U(1)$ current $T_1(z)$ becomes a primary field of spin 1 with respect to this Virasoro algebra.\\
We now turn our attention to the eigenvalues of $D_0(q)$,in terms of the number operators we can write $D_0(q)$ as:
\be D_0(q)=q^{-1}\xi_0\eta_0+q^{-1}\sum_{n=1}^\infty(q^{-2n}N^+_n-q^{2n}N^-_n)  \ee
from which it is straightforward to obtain the eigenvalues:
\be  D_0(q)\big|\mathcal{D}\big>=q^{-1}(\sum_{n\in\mathcal{P}_1} q^{-2n}-\sum_{n\in\mathcal{P}_2}q^{2n})\big|\mathcal{D}\big> \ee
where again we have used the notation $\mathcal{D}=(\mathcal{P}_1|\mathcal{P}_2)$. \\
We now consider the expansion
\be D_0(q)=\sum_{k=0}^\infty \hbar^k D_0^{(k)}  \ee
where
\be D_0^{(k)} =\frac{(-1)^k}{k!}\big(\xi_0\eta_0+\sum_{s=0}^k\binom{k}{s}2^s I_s) \ee
where once more the $I_s$ are the integrals of motion.\\
We now turn our attention to the possibility of forming a representation of $\hat{\mathcal{D}}_q$ which has a nontrivial logarithmic structure, it tuns out that this is possible. We start by observing that if we add a total derivative of the $\eta$ field to the stress energy tensor:
\be  \tilde{T}_2(z)=T_2(z)+\delta \partial\eta(z)  \ee
The central charge remains the same, in this case, however the new generator of dilatations has nontrivial Jordan block structure:
\be  \tilde{L}_0\Omega=0 \ee
\be  \tilde{L}_0\omega=\Omega   \ee
from this inspiration one notices that if one defines:
\be  \tilde{D}(q,z)=D(q,z)-2 \eta(q z)  \ee
it is straightforward to check that this current satisfies the same commutation relations of the old one, however the $\tilde{D}_0(q)$ generator has the nontrivial Jordan block structure:
\be  \tilde{D}_0(q)\Omega=0 \ee
\be  \tilde{D}_0\omega=-2q^{-1}\Omega   \ee

\section{Characters}
We start this section by computing the following character:
\be \chi(q_0,q_1)=\textrm{Tr}(q_1^{I_1}q_0^{I_0})  \ee
where the trace is over all the eigenstates of the IOM in the vacuum sector, this can be expressed as the 
following sum over two column diagrams:
\be  \chi(q_0,q_1)=\sum_{\mathcal{D}} q_1^{I_1(\mathcal{D})}q_0^{I_0(\mathcal{D})} \ee
if we consider this sum over all two column diagrams of maximal height $L$, that is the length of each partition $\mathcal{P}_1$, $\mathcal{P}_2$ cannot exceed the value $L$, and the heights inside the partitions are also less than $L$, we have the following finitized character:
\be   \chi^{(L)}(q_0,q_1)= \prod_{n=1}^L (1+q_1^n q_0)(1+q_1^n q_0^{-1})    \ee
in the limit $L\to\infty$ we obtain the well known $sl(2)\otimes \textrm{Vir}$ character (in the Noveau-Schwartz sector) \cite{kausch2}:
\be   \chi(q,w)= \prod_{n=1}^\infty (1+q^n w)(1+q^n w^{-1})    \ee
The calculation can be repeated in the twisted sector (Ramond) and one obtains the result:
\be   \chi(q,w)= \prod_{n=1}^\infty (1+q^{n-\frac{1}{2}} w)(1+q^{n-\frac{1}{2}} w^{-1})    \ee
We now turn our attention to  of $I_2$, which is also $W^{(0)}_0$, the generator of the $\mathcal{W}_3$ algebra, if we consider the following character:
\be \chi(q_0,q_1,q_2)=\mathcal{N}(q_0,q_2)\textrm{Tr}(q_2^{I_2}q_1^{I_1}q_0^{I_0})  \ee
we have the following finitization in the Noveau-Schwartz (NS) sector :
\be \chi^{(L)}(q_0,q_1,q_2)=\mathcal{N}(q_0,q_2)\prod_{n=1}^\infty (1+q_2^{n^2}q_1^n q_0)(1+q_2^{-n^2}q_1^n q_0^{-1})  \ee
whereas in the Ramond (R) sector:
\be \chi^{(L)}(q_0,q_1,q_2)=\mathcal{N}(q_0,q_2)\prod_{n=1}^\infty (1+q_2^{(n-\frac{1}{2})^2}q_1^{n-\frac{1}{2}} q_0)(1+q_2^{-(n-\frac{1}{2})^2}q_1^{n-\frac{1}{2}} q_0^{-1})  \ee
where we have inserted a normalization factor $\mathcal{N}(q_0,q_2)$ to make the product convergent in the $L\to\infty$ limit, this normalization looks like in the NS sector:
\be  \mathcal{N}(q_0,q_2)=q_0^{\sum_{n=1}^L n^0} q_2^{\sum_{n=1}^L n^2} \ee
whereas in the R sector:
\be   \mathcal{N}(q_0,q_2)=q_0^{\sum_{n=1}^L (n-\frac{1}{2})^0} q_2^{\sum_{n=1}^L (n-\frac{1}{2})^2}  \ee
If we choose to use $\zeta$ function regularization for the normalization factors and use the fact that the $\zeta$ function vanishes at negative even integers, we can replace the normalization factors by $1$ in the $L=\infty$ limit so that in this limit we obtain the $\mathcal{W}_3$ character of \cite{characterW3}, which has been proven in \cite{KZ} to be related to the generating function counting maps of curves of genus $g > 1$ to a curve of genus 1.\\
We now make a leap forward and define in complete generality:
\be  \chi^{(L)}(q_0,q_1,q_2,\ldots)= \mathcal{N}(q_0,q_2,\ldots)\textrm{Tr}(\prod_{i=0}^\infty q_i^{I_i})  \ee
which can be evaluated in the NS sector as:
\be  \chi^{(L)}(q_0,q_1,q_2,\ldots)= \mathcal{N}(q_0,q_2,\ldots)\prod_{n=1}^L(1+\prod_{i=0}^\infty q_i^{n^{i}})(1+\prod_{i=0}^\infty q_i^{(-1)^{i+1}n^{i}})  \ee
whereas in the R sector one has
\be  \chi^{(L)}(q_0,q_1,q_2,\ldots)= \mathcal{N}(q_0,q_2,\ldots)\prod_{n=1}^L(1+\prod_{i=0}^\infty q_i^{(n-\frac{1}{2})^{i}})(1+\prod_{i=0}^\infty q_i^{(-1)^{i+1}(n-\frac{1}{2})^{i}})  \ee
where the normalization in the NS sector is
\be   \mathcal{N}(q_0,q_2,\ldots)=q_0^{\sum_{n=1}^L n^0} q_2^{\sum_{n=1}^L n^2} q_4^{\sum_{n=1}^L n^4}\ldots  \ee
whereas in the R sector
\be   \mathcal{N}(q_0,q_2,\ldots)=q_0^{\sum_{n=1}^L (n-\frac{1}{2})^0} q_2^{\sum_{n=1}^L (n-\frac{1}{2})^2}  q_4^{\sum_{n=1}^L (n-\frac{1}{2})^4}\ldots  \ee
By using again the $\zeta$ function regularization we can replace these factors by $1$ in the $L=\infty$ limit.\\
These characters are almost with the characters projective representation of the algebra of differential operators on the circle \cite{SBAO} which is related to the counting function of the m-simple branched covers with a fixed genus of an elliptic curve \cite{HO}. This character is proven in \cite{SBAO} to be quasi modular of weight $0$. In order to make our characters coincide with those of \cite{SBAO}, and hence be quasi modular, we need to multiply them by the normalization factor:
\be  \tilde{\mathcal{N}}(q_1,q_3,q_5,\ldots)= q_1^{\zeta(-1,1-\frac{\delta}{2})} q_3^{\zeta(-3,1-\frac{\delta}{2})} q_5^{\zeta(-5,1-\frac{\delta}{2})}\ldots  \ee
where $\delta=0$ in the NS sector whereas $\delta=1$ in the R sector. Multiplying our character by this factor can be understood as shifting the $I_{2n-1}$ by a constant and thus defining:
\be  \mathcal{I}_{2n-1}=I_{2n-1}+\zeta(1-2n,1-\frac{\delta}{2}) \ee
This is very natural since the $I_{2n-1}$ coincide with the local integrals of motion of $\cite{blz}$ only upon subtracting the vacuum eigenvalue, which we have now restored, the $\mathcal{I}_{2n-1}$ are now really the local integrals of motion and their character:
\be  \chi(q_0,q_1,q_2,\ldots)=\textrm{Tr}(q_0^{I_0}q_1^{\mathcal{I}_1}q_2^{I_2}q_3^{\mathcal{I}_3}\ldots) \ee
coincides with the character of \cite{SBAO} and is now quasi modular of weight zero.\\
We now want to compute the finitized character of the $D_0(t)$ generator of the algebra $\hat{\mathcal{D}}_t$ which is obviously defined as:
\be \chi^{(L)}(D_0(t),q)= \sum_{\mathcal{D}} q^{D_0(t,\mathcal{D})} \  \ee
this can be evaluated to:
\be  \chi^{(L)}(D_0(t),q)=\prod_{n=1}^L q^{-t^{2n-1-\delta}}(1+q^{t^{-2n-1+\delta}})(1+q^{t^{2n-1-\delta}})  \ee
which in the $L\to\infty$ limit becomes:
\be  \chi(D_0(t),q)=q^{\frac{t^{1-\delta}}{t^2-1}}\prod_{n=1}^\infty (1+q^{t^{-2n-1+\delta}})(1+q^{t^{2n-1-\delta}})  \ee
Where $\delta=0$ in the NS sector and $\delta=1$ in the R sector.\\
We now recall that
\be D_0^{(k)}(\mathcal{D}) =\frac{(-1)^k}{k!}\sum_{s=0}^k\binom{k}{s}2^s I_s(\mathcal{D}) \ee
from this it is clear that we can compute the character of the $D_0^{(k)}$ in terms of $ \chi(q_0,q_1,q_2,\ldots)$ by setting:
\be  q_s=q^{\frac{(-1)^k 2^s}{k!}\binom{k}{s}}  \ee

\section{Conclusion}
In this paper we have introduced a new free field realization of the central extension of the Lie algebra of difference operators on the circle in terms of the $\eta$-$\xi$ ghost system. We have shown that this free field realization admits a nontrivial Jordan block structure for the deformed dilatation generator. We have also elucidated the relation of this algebra and of the $\mathcal{W}_{1+\infty}$ algebra to the Local Integrals of Motion of \cite{blz}. Furthermore we have computed the characters of all the integrals of motion and shown that they are related to the generating functions for the counting of the m-simple branched covers with a fixed genus of an elliptic curve. We have also computed the character of the deformed dilatation generator.

\section{Acknowledgements}
The author acknowledges financial support from Fondo Sociale Europeo (Regione Lombardia), through the grant ÒDote ricercaÓ.\\

\end{document}